\documentclass[reprint,showpacs,amsmath,amssymb,xcolor=dvipsnames,prl,longbibliography,superscriptaddress,floatfix]{revtex4-1}
\pdfoutput=1

\usepackage[T1]{fontenc}
\usepackage[pdftex]{graphicx}% Include figure files
\usepackage{bm}% bold math
\usepackage[pdftex,colorlinks=true,citecolor=blue,urlcolor=blue,linkcolor=black]{hyperref}
\usepackage{braket}
\usepackage[dvipsnames]{xcolor}
\usepackage[version-1-compatibility,binary-units,abbreviations=true]{siunitx}
\usepackage{comment}

\usepackage{physics}
\usepackage{mathrsfs}
\usepackage{xspace}
\usepackage{hyperref}

\newcommand{\oR}{\omega}

\definecolor{uc_blue}{HTML}{3891a6}
\definecolor{uc_red}{HTML}{c61a27}

\begin{document}

%%%%%%%%%%%%%%%%%%%%%%%%%%%%%%%%%%%%%%%%%%%%%%%%%%%%%%%%%%%%%%%%%%%%%%%%%%%%%%%%%
%						    	Title and Authors	    						%
%%%%%%%%%%%%%%%%%%%%%%%%%%%%%%%%%%%%%%%%%%%%%%%%%%%%%%%%%%%%%%%%%%%%%%%%%%%%%%%%%

\title{ Engineering single-atom angular momentum eigenstates in an optical tweezer}

\author{Philipp~Lunt}
    \email{lunt@physi.uni-heidelberg.de}
	\affiliation{Physikalisches Institut der Universit\"at Heidelberg, Im Neuenheimer Feld 226, 69120 Heidelberg, Germany}
\author{Paul~Hill}
	\affiliation{Physikalisches Institut der Universit\"at Heidelberg, Im Neuenheimer Feld 226, 69120 Heidelberg, Germany}
\author{Johannes~Reiter}
	\affiliation{Physikalisches Institut der Universit\"at Heidelberg, Im Neuenheimer Feld 226, 69120 Heidelberg, Germany}
\author{Philipp~M.~Preiss}
	\affiliation{Max Planck Institute of Quantum Optics, Hans-Kopfermann-Str. 1, 85748 Garching, Germany}
	\affiliation{Munich Center for Quantum Science and Technology (MCQST), Schellingstr. 4, 80799 München, Germany } 
\author{Maciej~Ga\l ka}
	\affiliation{Physikalisches Institut der Universit\"at Heidelberg, Im Neuenheimer Feld 226, 69120 Heidelberg, Germany}
\author{Selim~Jochim}
	\affiliation{Physikalisches Institut der Universit\"at Heidelberg, Im Neuenheimer Feld 226, 69120 Heidelberg, Germany}
\date{\today  }

%%%%%%%%%%%%%%%%%%%%%%%%%%%%%%%%%%%%%%%%%%%%%%%%%%%%%%%%%%%%%%%%%%%%%%%%%%%%%%%%%
%						        	Abstract        	   						%
%%%%%%%%%%%%%%%%%%%%%%%%%%%%%%%%%%%%%%%%%%%%%%%%%%%%%%%%%%%%%%%%%%%%%%%%%%%%%%%%%

\begin{abstract}
We engineer angular momentum eigenstates of a single atom by using a novel all-optical approach based on the
interference of Laguerre-Gaussian beams. We confirm the imprint of angular momentum by measuring the two-
dimensional density distribution and by performing Ramsey spectroscopy in a slightly anisotropic trap, which
additionally reveals the sense of rotation. This article provides the experimental details on the quantum state
control of angular momentum eigenstates reported in P. Lunt et al., \href{https://journals.aps.org/prl/abstract/10.1103/PhysRevLett.133.253401}{Phys. Rev. Lett. 133, 253401 (2024)}.
\end{abstract}
\maketitle

%%%%%%%%%%%%%%%%%%%%%%%%%%%%%%%%%%%%%%%%%%%%%%%%%%%%%%%%%%%%%%%%%%%%%%%%%%%%%%%%%
%						        	Main Text         	   						%
%%%%%%%%%%%%%%%%%%%%%%%%%%%%%%%%%%%%%%%%%%%%%%%%%%%%%%%%%%%%%%%%%%%%%%%%%%%%%%%%%

\section{Introduction}

Quantum state engineering at the level of individual constituents forms a cornerstone of modern quantum technologies, ranging from quantum metrology~\cite{Ye_2008, Pezze_RevModPhys.90.035005} to quantum simulation~\cite{Bloch_RevModPhys.80.885} and computation~\cite{Preskill_2018}.
It enabled breakthroughs in cooling the motional degree of freedom of nanoparticles~\cite{Deli__2020}, ions~\cite{Blatt_2012Natphys}, neutral atoms~\cite{Kaufman_PhysRevX.2.041014, Thompson_PhysRevLett.110.133001, Serwane_2011, brown2022timeofflight}, and even molecules~\cite{Anderegg_doi:10.1126/science.aax1265, Bao_PhysRevX.14.031002}.
These platforms are particularly versatile as they offer precise geometric shaping of arbitrary optical trapping potentials~\cite{Gross_2021, Navon_2021, Weitenberg_2021, Browaeys_2020}, thereby facilitating detailed control over the quantum state.

The manipulation of quantum systems makes use of energy and momentum transfer when light interacts with matter. Specific light fields such as Laguerre-Gaussian (LG) beams carry well-defined quanta of orbital angular momentum $l \hbar$~\cite{Allen_PhysRevA.45.8185} (in addition to their intrinsic angular momentum determined by their polarization) and can induce a mechanical rotation in matter~\cite{Andrews_Babiker_2012}. The transfer of orbital angular momentum from an LG beam to a macroscopic nanoparticle demonstrated the ability of light fields to exert torque~\cite{He_PhysRevLett.75.826}. Furthermore, the transfer of orbital angular momentum to a macroscopic quantum state forming a Bose Einstein condensate showed the quantization of the angular momentum transfer~\cite{Andersen_PhysRevLett.97.170406}. However, the angular momentum control of a single neutral atom has remained elusive.

\begin{figure}
    \centering
	\includegraphics{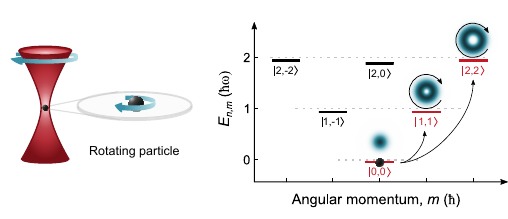}
    \caption{
    {Engineering angular momentum eigenstates.}  
    A single atom, prepared in the ground state of the optical tweezer is set into rotation by the rotating external light field. The Gaussian trapping potential forms approximately a 2D harmonic oscillator with trap frequency $\oR$. We label the states $\ket{n,m}$ by shell number $n$ and angular momentum number $m$; for the states with maximal angular momentum in each shell $n=m$ we show the state's density profiles, where the arrows indicate the phase winding. The rotating light field selectively couples  the ground state to non-zero angular momentum states (black dashed arrows).
    \label{fig:experiment}
    }
\end{figure}

\begin{figure}
    \centering
	\includegraphics{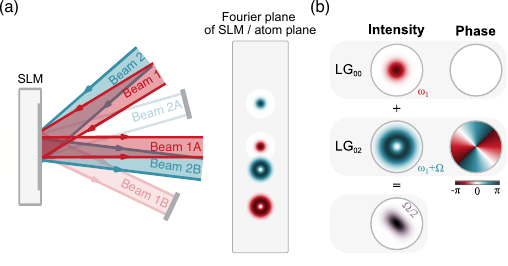}
    \caption{
    {Experimental setup to generate rotating optical potentials.}  
    (a) A spatial light modulator (SLM) is placed in the Fourier plane of the atom. By displaying an appropriate phase pattern on the SLM, we generate two outgoing beams (A and B) from a single incident beam with a different phase winding. Hence, two incident beams, Beam 1 and 2, with a small relative angle with respect to each other generate four outgoing beams. The SLM is used to overlap Beam 1A forming the optical tweezer and Beam 2B forming the perturbation, here displaced for clarity.
    (b) The tweezer is formed by a radially symmetric Gaussian beam. We imprint a phase winding on the perturbation beam to generate an LG mode of order $l$ (here $l=2$) in the atomic plane. The interference of the two beams creates an $l-$fold symmetric trap (for $l=2$ an elliptically shaped trap). The relative frequency $\Omega $ between the interfering beams and the order $l$ of the LG mode sets the rotation frequency $\Omega / l$ of the deformed potential. The optical frequency of the first beam is denoted $\omega_1$.
    \label{fig:slm}
    }
\end{figure}

In this article, we present an all-optical approach for injecting orbital angular momentum to a single atom by rotating an anisotropy of an optical tweezer. The precise control of the shape of the light field and its rotational speed, together with the small anharmonicity of the potential, enables us to selectively address motional states that differ in angular momentum and energy. In  Fig.~\ref{fig:experiment}, we illustrate this process and show the in-plane tweezer potential, which approximately forms a two-dimensional (2D) harmonic potential with trap frequency $\oR$. This work introduces a novel tool for quantum state engineering and lays the groundwork for studies on interacting many-body systems in rapidly rotating optical traps~\cite{Lunt2024}.

\section{Rotating optical potentials}

Our approach to create rotating optical traps is based on the interference of two Laguerre-Gaussian beams with waist $W$ and electric field $\text{LG} _{0l} (r, \varphi)  \propto (r/W)^{|l|} e^{i l \varphi } e^{-r^2 / W^2}$, where $r,\varphi$ represent the polar coordinates. The main trap of the atoms is formed by a Gaussian $\text{LG} _{00}$ beam which is then interfered with a second $\text{LG} _{0l}$ beam to induce rotation. The LG$_{0l}$ mode carries $l \hbar$ quanta of orbital angular momentum which is incorporated in the phase winding $e^{i l \varphi }$ that breaks the rotational symmetry of the combined in-plane intensity pattern. Furthermore, by modulating the relative phase between both beams via the angular frequency detuning $\Omega$ we can engineer the time dependent intensity distribution (see also Appendix A)
\begin{equation}
    \begin{split}
    &I(r, \varphi) = \left|\sqrt{P}\text{LG}_{00}-\sqrt{P_l}e^{-i\Omega t}\text{LG}_{0l}\right|^2\\
    &\sim (1 - \beta^*_lr^l\cos(l\varphi - \Omega t))e^{-2r^2/W^2}.
\end{split}
\end{equation}
Here, $P$, and $P_l$ denote the power of the main tweezer and the perturbation beam, respectively, while $\beta_l^*$ denotes the resulting strength of the interference term.

All light fields are formed via a spatial light modulator (SLM) in the Fourier plane of the atoms. In order to reduce optical phase aberrations stemming from the optical elements in the beam path such as the objective and the vacuum window, we measure the optical aberrations directly with the atoms via a phase-shifting interferometry algorithm \cite{hill2024optical, Zupancic_2016}.

A sketch of the SLM setup to generate rotating optical potentials is illustrated in Fig.~\ref{fig:slm}(a). Two Gaussian laser beams (Beam 1 in red, Beam 2 in blue) are superimposed on the SLM under a non-zero relative angle. The phase pattern on the SLM forms two outgoing beams per incident beam~\cite{Liesener_2000}, denoted A and B. On Beams A we choose a constant phase profile, leading to a Gaussian beam LG$_{00}$ in the Fourier plane, while on Beams B we imprint the $2\pi l$ phase winding, which approximately forms an $l$th-order LG mode $\text{LG}_{0l}$ in the Fourier plane~\footnote{The phase winding imprinted via the SLM enforces the outgoing beam to have a intensity depletion at its center. However, the exact radial distribution will deviate from that of an $\text{LG}_{0l}$ mode. In general, the beam will be a superposition of $\text{LG}_{kl}$ modes with differing $k$ but fixed $l$.}.  The relative angle between the outgoing beams is adjusted with the SLM such that two of the four beams are overlapped (Beam 1A and Beam 2B in Fig.~\ref{fig:slm}). The other unwanted diffraction orders are spatially filtered with an iris in a Fourier plane behind the SLM, which is imaged on the plane of the atoms.
Beam 1A constitutes the main optical tweezer, while beam 2B is the perturbation. In this configuration the SLM allows to independently modify the local phase of the two overlapping outgoing beams. As both beams originate from different beams incident on the SLM properties like beam power and global phase are individually addressable as well.

\begin{figure}
    \centering
	\includegraphics{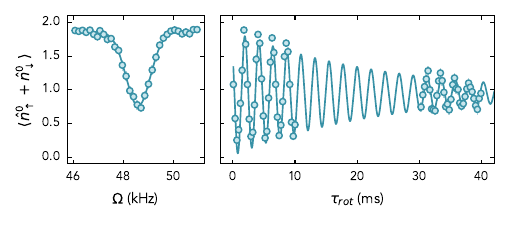}
    \caption{
    {Resonance spectrum and Rabi oscillations.}  
    (a) Resonance spectrum of the excitation from the ground state $\ket{0,0}$ to the state $\ket{2,2}$. The resonance is shifted down to $\Omega _\text{res} \approx \SI{1.73}{}\oR$ compared to $2\oR$ due to the anharmonicity of the optical potential. 
    (b) Rabi oscillations between the ground state $\ket{0,0}$ and state $\ket{2,2}$ with Rabi rates $\Omega _\text{rabi} / 2 \pi \approx \SI{0.44}{kHz}$ and a coherence time $\tau _\text{coh} = \SI{23(1)}{ms}$.
    \label{fig:rabi}
    }
\end{figure}

The speed of rotation is set by the relative angular frequency $\Omega  $ of the tweezer and the perturbation beam, controlled via an acousto-optical modulator for each beam. This allows us to drive arbitrary frequency ramps, including a smooth increase of the rotation frequency or jumps. The optical trap geometrically rotates at rate $\Omega / l$, which is $l$-times slower than the frequency detuning between the beams. This reflects the $l$-fold symmetry of the LG$_{0l}$ mode arising from its phase winding $2\pi l$. In Fig.~\ref{fig:slm}(b),  we show as an example the case of an LG$_{02}$ mode which results in the elliptical shape of the trapping potential rotating with a rotation frequency ${\Omega / 2 }$.

\section{Coherent control of angular momentum states}

\begin{figure*}[t]
    \centering
	\includegraphics{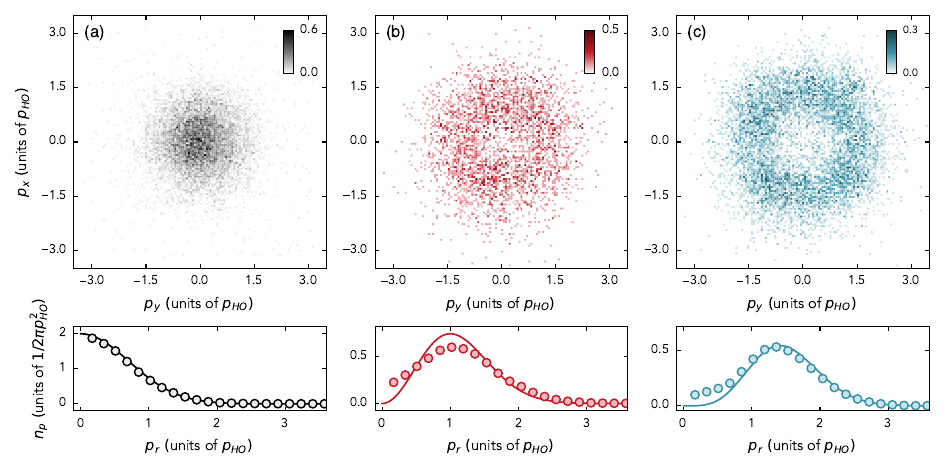}
    \caption{
    {Single atom angular momentum eigenstates.}  
    Normalized two-dimensional momentum-space density distribution (top), and azimuthally averaged radial density (bottom) of the ground state $\ket{0,0}$ with zero angular momentum (a), and states with non-zero angular momentum $\ket{1,1}$ (b), and $\ket{2,2}$ (c). The solid lines in the bottom row are theoretical calculations without free parameters. 
    \label{fig:densities}
    }
\end{figure*}

The experiment starts by loading a gas of  $^6$Li atoms from a magneto-optical trap into a red-detuned, crossed optical dipole trap. After a sequence of radio-frequency pulses, we end up with a balanced two-component mixture of $^6$Li in the hyperfine states $|F=1/2,m_\text{F}=1/2 \rangle $ and ${|F=3/2,m_\text{F}=-3/2 \rangle}$; in~\cite{Lunt2024} they are referred to as spin up $|\uparrow \rangle$ and spin down $ |\downarrow \rangle$, respectively. Next, we load the atoms from the crossed optical dipole trap into a tightly focused, cigar-shaped optical tweezer. We evaporate in the tweezer within $\SI{40}{ms}$ and reach a highly degenerate sample of roughly \SI{200}{} atoms in total after evaporation. Subsequently, we use the spilling technique pioneered in \cite{Serwane_2011} to prepare one spin up and one down atom in the ground state of the optical tweezer with fidelities $ \SI{95(3)}{\%}$; the spilling procedure is performed at \SI{300}{G}. We ramp the magnetic field to $\SI{568}{G}$ at which the spin states are non-interacting, which allows us to effectively consider a single atom in the ground state throughout the remaining paper (the other atom acts as an identical copy of the first one).

Our optical tweezer is formed by a Gaussian beam with waist $W \approx \SI{1.1}{\micro m}$ and leads to an approximately harmonic potential with radial and axial trap frequency $\oR / 2 \pi \approx \SI{28.1}{kHz} $ and $\omega _\text{ax} / 2\pi \approx \SI{3.7}{kHz}$, respectively. Since we prepare a single atom in the ground state and the rotation only couples to the radial motion of the atom, we neglect the axial degree of freedom. The in-plane potential in the harmonic expansion reads

\begin{align}
    V_\text{pot} = -V_0 e^{-2r^2/W^2}  \sim \frac{ m_\text{Li} \oR ^2}{2}  r^2  + \mathcal{O}(r^4),
    \label{eq:potential}
\end{align}

where $V_0$ denotes the potential depth of the tweezer, $m_\text{Li}$ is the mass of $^6$Li, and the constant energy offset is neglected on the right-hand side of the equation. Corrections to the harmonic potential can be treated perturbatively, and we find that level shifts for the states interesting to the present work are on the order of several kHz (see Appendix B). We emphasize that the anharmonicity breaks the equidistance of the level spacing in the harmonic trap which enables closed transitions between two distinct states.

To reach states with non-zero angular momentum we use our all-optical approach described in the preceding section. Low-lying states in the trap (i.e. states that do not probe the Gaussian envelope of the perturbation beam) experience the rotating perturbation (see also Appendix A)

\begin{align}
    V_\text{p} = \beta _l r^l  \left( e^{i l \varphi } e^{-i\Omega t} + H.c \right), \label{eq:vpert}
\end{align}

for an LG$_{0l}$ mode, and perturbation strength $\beta _l=V_0\beta_l^*/2 \ll 1$. The perturbation couples the ground state $\ket{0,0}$ to an angular momentum state $\ket{n, m=l}$; the coupling is resonant when $\hbar\Omega$ is equal to the energy difference between the states $E_{nm} - E_{00}$. 
To selectively address individual motional states we make use of the anharmonicity to render the ground state $\ket{0,0}$ and the state $\ket{n,m}$ a two-level system, in case the Rabi rate remains lower than the anharmonicity.

\begin{figure*}
    \centering
	\includegraphics{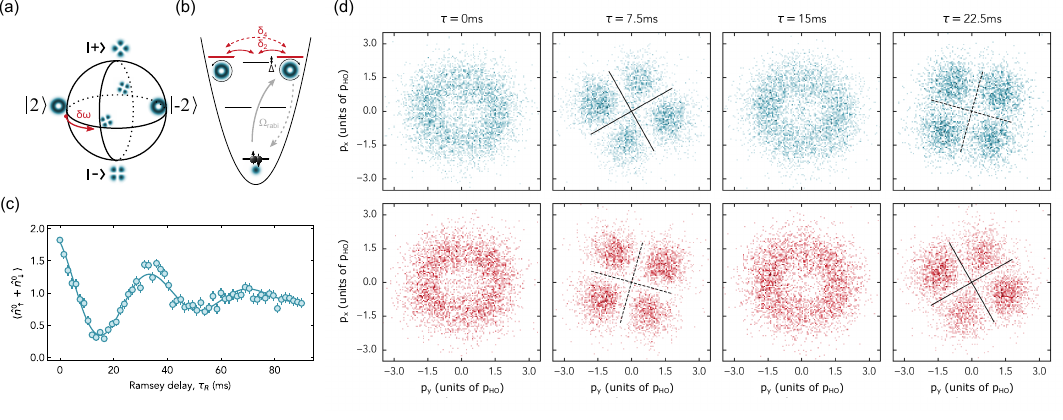}
    \caption{
    {Ramsey spectroscopy.}  
    (a) In the presence of anisotropy the superposition states $|\pm \rangle = 1/\sqrt{2} (|2,2\rangle \pm |2,-2\rangle ) $ form the new eigenstates of the system with an energy difference given by the anisotropy $\delta \omega $. The initialized state $|2,2 \rangle $ is an equal superposition of the eigenstates $|+\rangle , |- \rangle $ and hence evolves over time $\tau _\text{R}$ on the equator of the Bloch sphere.
    (b) The anisotropy couples states which differ in angular momentum. We harness this effect to perform Ramsey spectroscopy by preparing the state $\ket{2,2}$ via a $\pi-$pulse, followed by a Ramsey delay time $\tau _\text{R}$ and a second $\pi -$pulse to de-excite the atoms to the ground state. Depending on the contribution of the $|2,-2\rangle$ state, the overlap with the ground state oscillates.  
    (c) Ramsey spectrum of the state $\ket{2,2}$. We measure an anisotropy of $\delta \omega / 2 \pi= \SI{27.3(4)}{Hz}$ at a trap frequency $\oR / 2 \pi \approx \SI{28.1}{kHz}$ yielding a relative anisotropy $\delta \oR / \oR = \SI{9.6e-4}{}$. 
    The coherence time is $\tau ^\text{(non-int)} _\text{coh} = \SI{35(3)}{ms}$. The amplitude oscillates between two and zero atoms as we have two identical (non-interacting) copies of the same state in the tweezer. 
    d) Normalized density distribution of clockwise (upper row) and counterclockwise (lower row) evolving state at different evolution times.
    \label{fig:ramsey_spectrum}
    }
\end{figure*}

To demonstrate our exquisite control of preparing motional quantum states we drive Rabi oscillations between $\ket{0,0}$ and $\ket{2,2}$. To determine the resonance frequency we spectroscopically measure the single-particle occupation number in the ground state $\langle\hat{n}^{0}_{\uparrow}+\hat{n}^{0}_{\downarrow}\rangle $ after applying a rotating perturbation for $\tau = \SI{350}{\mu s}$ at different rotation frequencies, shown in Fig.~\ref{fig:rabi}(a); here $\hat{n}^{0}_{\uparrow(\downarrow)}$ is the number operator for the spin up (down) particle in the ground state. We measure a resonance frequency $\Omega / 2\pi \approx \SI{48.59}{kHz}$ which corresponds to $\SI{1.73}{\oR}$ in units of the radial trap frequency $\oR$. The resonance frequency is downshifted from $\SI{2}{\oR}$ due to the anharmonicity of the optical potential. Indeed, the observed frequency agrees well with the expected resonance based on first order perturbation theory which gives $\SI{1.75}{\oR}$ (c.f. Appendix B). On resonance, we drive coherent oscillations between the ground state $\ket{0,0}$ and the state $\ket{2,2}$, see Fig.~\ref{fig:rabi}(b), with a Rabi rate $\Omega _\text{rabi} / 2 \pi \approx \SI{0.44}{kHz}$, and a coherence time $\tau _\text{coh} = \SI{23(1)}{ms}$ significantly longer than the duration of a $\pi -$pulse.

The precise control of angular momentum eigenstates requires to overcome the following experimental challenges. 
First, the relative position of the optical tweezer and the perturbation is required to be aligned on the order of the radial extent of the wavefunction, which in our case is typically around $\SI{200}{nm}$. To this end, we use an LG$_{00}$ mode as a resonant perturbation whose position is scanned across the two-dimensional atom plane. By measuring the atom loss we determine the relative position in the atom plane to $\sim \SI{100}{nm}$. Second, the anisotropy $\delta \oR$ of the optical tweezer breaks the rotational symmetry of the system and sets an upper time scale $1/\delta \oR$ for the preparation of the angular momentum eigenstates (see next section for details).

After preparing an angular momentum eigenstate, we release the atom from the tweezer and perform a time-of-flight expansion for $t_\text{tof} = \SI{2.5}{ms}$ to measure the momentum of the atom using our single atom fluorescence imaging technique~\cite{Bergschneider_PhysRevA.97.063613}. In order to keep the atom within the depth of focus of our objective during the expansion, we rapidly turn on a 2D lattice with an axial confinement approximately matching the axial trap frequency of the optical tweezer. We note that this time-of-flight expansion is self-similar, reflecting the fact that the harmonic oscillator wavefunctions have the same shape in their position and momentum space representation. 
We reconstruct the 2D momentum density distribution of the first three angular momentum eigenstates $\ket{0,0},\ket{1,1},\ket{2,2}$ by taking $\SI{10047}{}, \SI{3398}{}, \SI{7998}{}$ snapshots of the wavefunction, shown in Fig.~\ref{fig:densities}(a,b,c), respectively. While all densities exhibit a rotationally symmetric distribution, the non-zero angular momentum states show the expected density depletion at zero momenta and a maximum at $\sqrt{m}p_\text{HO}$, with the momentum scale in harmonic oscillator units being $p_\text{HO} = \sqrt{\hbar m_\text{Li}\omega}$.

To quantitatively compare the measured distribution to the eigenstates of the 2D harmonic oscillator, we determine the radial densities $n_\text{p}$ by azimuthally averaging over the obtained 2D densities, shown in the lower row of Fig.~\ref{fig:densities} for the respective angular momentum state $\ket{m,m}$. We find good agreement between the fit-free theoretical curve and our experimental data. The largest deviations occur at small momenta which we attribute to an imperfect $\pi -$pulse excitation caused by fluctuations of the trap frequency (below $\SI{1}{\percent}$), which leaves the atom in the ground state and therefore contributes to the density at zero momenta.

\section{Time evolution of angular momentum states }

To confirm that the central depletion of the density distribution stems from a phase winding given by the imprint of angular momentum, we investigate the time evolution of the state $\ket{2,2}$ in an anisotropic potential. In a radially symmetric trap an angular momentum state is an eigenstate of the Hamiltonian. However, in an anisotropic potential the angular momentum states become a superposition of the true eigenstates of the system and thus evolve in time. We make use of a small residual anisotropy present in the optical potential which we attribute to a slight ellipticity of the tweezer. The new effective energy eigenstates of this system are then formed by $\ket{\pm} = 1/\sqrt{2}\left(\ket{2,2} \pm \ket{2,-2}\right)$ with an energy difference defined as the anisotropy $\delta \omega$ (for details see Appendix C). The states $|\pm \rangle$ form a two-level system,  which can be depicted on the Bloch sphere, shown in Fig.~\ref{fig:ramsey_spectrum}(a). Therefore, for the $\ket{2,2}$ state we expect a characteristic time evolution oscillating between the $|2,2\rangle $ and $|2,-2\rangle $ states.

We perform Ramsey spectroscopy on the state $\ket{2,2}$ and observe coherent oscillations with a frequency given by the anisotropy $\delta \oR$. In Fig.~\ref{fig:ramsey_spectrum}(b) we outline the experimental protocol. We use a $\pi-$pulse to inject $2\hbar$ quanta of angular momentum (gray solid arrow). Subsequently, we let the system evolve for a delay time $\tau _\text{R}$ (red arrows), after which we use a second $\pi-$pulse to de-excite the evolved state to the ground state (gray dashed arrow). We measure the single-particle occupation number in the ground state $\langle\hat{n}^{0}_{\uparrow}+\hat{n}^{0}_{\downarrow}\rangle $ in Fig~\ref{fig:ramsey_spectrum}(c), which oscillates with the energy difference given by the anisotropy $\delta \oR / 2\pi = \SI{27.3(4)}{Hz}$. This yields a relative anisotropy $\delta \oR / \oR = \SI{9.6e-4}{}$. The state is evolving on a time scale much longer than the duration of the $\pi$-pulse, which sets the time scale on which we prepare and detect the state $\ket{2,2}$. The coherence time of the Ramsey oscillations is $\tau ^\text{(non-int)} _\text{coh} = \SI{35(3)}{ms}$. 
We argue that this timescale is limited by noise of experimental parameters, predominantly the trap depth, leading to loss of coherence via coupling to other energy levels in the trap. We expect the coherence time to be strongly dependent on the noise spectrum and levels close to the targeted state. Indeed, we observe a significant increase in coherence time when energy eigenstates close to the prepared state are shifted away, as demonstrated in the accompanying letter~\cite{Lunt2024}.

In Fig.~\ref{fig:ramsey_spectrum}(d), we show the 2D density distribution after preparing the $\ket{2,2}$ state and letting it evolve in the slightly anisotropic optical potential for different quarter periods of the Ramsey delay time. The density evolves from the $\ket{2,2}$ state at $\tau = 0$, to an equal superposition of the $\ket{2, \pm 2}$ states at $\tau = T/4$, it continues to the $\ket{2,-2}$ state at $\tau = 2T/4$ and further evolves again to a superposition of the $\ket{2,\pm 2}$ states at $\tau = 3T/4$, however, now tilted by \SI{45}{\degree} with respect to the state at $\tau = T/4$ (dashed black cross). 
By reversing the phase winding on the SLM we prepare the $\ket{2,-2}$ state, which starts the precession on the Bloch sphere from another starting point. Thereby, it confirms the imprint of the expected phase winding of $\pm 2\times 2\pi $, which corresponds to an angular momentum of $2\hbar$.

We further observe a slight deviation from the expected densities at $\tau = T/4, ~3T/4$ (non-vanishing center density connecting one pair of lobes diagonally), which we attribute to a weak coherent admixture of the $\ket{2, 0}$ state that mediates the effective time evolution between the $\ket{2,\pm 2}$ states (c.f. Appendix C).

\section{Conclusion}

We have demonstrated motional control of angular momentum eigenstates of a single atom in an optical tweezer. By interfering the optical tweezer with an LG beam of order $l=m$ we coherently coupled the ground state to a non-zero angular momentum eigenstate $\ket{m,m}$. We confirmed the preparation of the angular momentum eigenstates by measuring the 2D densities and by observing the evolution of the state in a slightly anisotropic trap. The latter allowed us to reveal the phase imprint on the wavefunction through density measurements at different times.
Beyond the control of the single-particle angular momentum eigenstates, this technique enables the studies of ultracold atoms in optical potentials subjected to synthetic magnetic fields~\cite{Cooper_2008}, including the recent realization of a two-particle Laughlin state~\cite{Lunt2024}.

%%%%%%%%%%%%%%%%%%%%%%%%%%%%%%%%%%%%%%%%%%%%%%%%%%%%%%%%%%%%%%%%%%%%%%%%%%%%%%%%%
%					     Acknowledgements and Contributions   		   			%
%%%%%%%%%%%%%%%%%%%%%%%%%%%%%%%%%%%%%%%%%%%%%%%%%%%%%%%%%%%%%%%%%%%%%%%%%%%%%%%%%

\paragraph*{Funding}
This work has been supported by the Heidelberg Center for Quantum Dynamics, the DFG Collaborative Research Centre SFB 1225 (ISOQUANT), Germany’s Excellence Strategy EXC2181/1-390900948 (Heidelberg Excellence Cluster STRUCTURES) and the European Union’s Horizon 2020 research and innovation program under grant agreements No.~817482 (PASQuanS),  No.~725636 (ERC QuStA) and No. 948240 (ERC UniRand). This work has been partially financed by the Baden-Württemberg Stiftung.

\appendix
\section{Appendix A: Interference of Laguerre-Gaussian beams}
\label{Ap:LG}
The electric field of a Laguerre-Gaussian beam of order $0$, $l$ and waist $W$ at its focus is given by
\begin{equation}
    \text{LG}_{0l}(r, \varphi, z=0) = \sqrt{\frac{2}{\pi l! W^2}}\left(\frac{\sqrt{2}r}{W}\right)^le^{il\varphi}e^{-r^2/W^2}.
\end{equation}
To create a rotating optical potential we interfere the light beam of the main tweezer, a Gaussian $\text{LG}_{00}$-mode with power $P$, with a Laguerre-Gaussian beam $\text{LG}_{0l}$-mode with power $P_l$, resulting in the intensity distribution
\begin{equation}
\begin{split}
    &I(r, \varphi) = \left|\sqrt{P}\text{LG}_{00}-\sqrt{P_l}\text{LG}_{0l}\right|^2\\
    &\quad= \frac{2P}{\pi W^2}e^{-2r^2/W^2}\left|1 - \sqrt{\frac{P_l}{P}}\left(\frac{\sqrt{2}r}{W}\right)^le^{il\varphi}\right|^2.
\end{split}
\end{equation}
High intensity seeking atoms experience the potential
\begin{equation}
\begin{split}
    &V(r, \phi) = -V_0e^{-2r^2/W^2}\\
    &+\beta_lr^l (e^{il\varphi}+e^{-il\varphi})e^{-2r^2/W^2} - \frac{1}{V_0}\beta_l^2r^{2l}e^{-2r^2/W^2},
\end{split}
\end{equation}
via the optical dipole force, where $V_0 = \gamma P/W^2$, $\gamma \approx 800 ~h\cdot\text{kHz µm}^2 /\text{mW}$, and $\beta_l = \gamma\sqrt{2^lP P_l}/W^{l+2}$. Here, the first term gives rise to a harmonic confinement at first order, with trapping frequency $\omega = 2W^{-2}\sqrt{\gamma P / m_\text{Li}}$ and length scale $l_0 = \sqrt{\hbar/m_\text{Li}\omega}$. As usual we are interested in the regime in which the harmonic approximation applies, i.e. $l_0\ll W$, and further in which the harmonic confinement is dominant, i.e. $\beta_ll_0^l/\hbar\omega\ll 1$. Additionally, we assume $\sqrt{2^l P_l/P}\left(l_0/W\right)^l\ll 1$ such that the $\beta^2$-term may be neglected. Lastly, by coherently altering the phase of the Laguerre-Gaussian beam according to $e^{-i\Omega t}$ we obtain Eq. \eqref{eq:vpert} from the main text,
\begin{equation}
    V_p = \beta_lr^l(e^{il\varphi - i\Omega t} + c.c.).
\end{equation}
Note, that Rabi rates between the harmonic oscillator states introduced by the rotating perturbation are on the order of $\beta_ll_0^l/\hbar$.

\section{Appendix B: Anharmonic level shifts}
The Gaussian potential in harmonic oscillator units can be rewritten according to
\begin{equation}
    V = \frac{1}{2}r^2 - \frac{1}{2g}(e^{-gr^2}-1)-\frac{1}{2}r^2 \equiv \frac{1}{2}r^2 + V_{anh},
\end{equation}
where $g = 2l_0^2/W^2 \approx 0.099$ in our case. We are primarily interested in states $\ket{m,m}$ of the harmonic oscillator, which in real space are described by the wavefunction $\varphi_{m, m}(r, \phi) = \sqrt{\frac{1}{\pi m!}}r^me^{im\phi}e^{-r^2/2}$. In first order perturbation theory such states experience an energy shift of
\begin{equation}
\begin{split}
    \Delta V_\text{anh} &= \frac{1}{2g}\left(1-\frac{1}{(1+g)^{m+1}}\right)-\frac{1}{2}(m+1) \\ \label{eq:anh_shift}
    &\approx - \frac{1}{4}(m+2)(m+1)g,
\end{split}
\end{equation}
where $\bra{m,m}e^{-gr^2}\ket{m,m}=(1+g)^{-(m+1)}$ and $\bra{m,m}r^2\ket{m,m}=m+1$ was used. We note that the expansion in $g$ of Eq. \eqref{eq:anh_shift} is equivalent to the expansion of the Gaussian potential in orders of $r^2$. We find that the first order in $g$ is sufficient for the states considered in this work.

\section{Appendix C: Anisotropic coupling}

\begin{figure}
    \centering
	\includegraphics{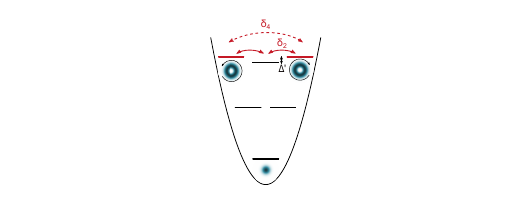}
    \caption{
    {Anisotropic coupling.}  
    The anisotropy breaks the azimuthal symmetry of the optical tweezer, leading to a coupling of states with $\Delta m = \pm 2 $ to first order and $\Delta m = \pm 4$ to second order in $\epsilon$ (see text).
    \label{fig:anisotropy}
    }
\end{figure}

We provide details on the effective coupling of the state $\ket{2,2}$ due to the presence of the remaining ellipticity of the optical tweezer, illustrated in Fig.~\ref{fig:anisotropy}. We can incorporate this ellipticity into the Gaussian trap model by inserting a factor $e^{-2\epsilon(x^2-y^2)/W^2}$ into the potential Eq.~(\ref{eq:potential}) which breaks the azimuthal symmetry. At first order in $\epsilon$ it introduces a perturbation term $\propto(r^2 e^{i2\varphi} + h.c.)$ that couples states with $\Delta m = \pm 2$ with a generic state-dependent coupling $\delta_{2}$ linear in $\epsilon$. Similarly, a term $\propto(r^4 e^{i4\varphi} + h.c.)$ appears at second order in $\epsilon$ (neglecting isotropic terms), and couples states with $\Delta m = \pm 4$ with coupling $\delta_{4}$. Therefore, at first order, states of the $\pm 2\hbar$ angular momentum manifold are coupled to states with $0\hbar$ quanta of angular momentum, and only at second order, there is direct coupling between states in the $\pm 2\hbar$ angular momentum manifold. Due to the anharmonicity of the optical trap the accessible $\ket{2,0}$ state is detuned from the $\ket{2,2}$,  $\ket{2,-2}$ states by $\Delta'\gg\delta_2$ which is on the order of kHz.
Hence, we expect an effective coupling $\delta_4-\delta_2^2/\Delta'\sim\epsilon^2$ only between the degenerate clockwise and counter-clockwise rotating states. The new effective energy eigenstates are given by $\ket{\pm} = 1/\sqrt{2}\left(\ket{2,2} \pm \ket{2,-2}\right)$. The energy difference between the eigenstates is defined as the anisotropy $\delta \omega \approx 2\delta_4-2\delta_2^2/\Delta'$. The states $|\pm \rangle$ then effectively form a two-level system.

%%%%%%%%%%%%%%%%%%%%%%%%%%%%%%%%%%%%%%%%%%%%%%%%%%%%%%%%%%%%%%%%%%%%%%%%%%%%%%%%%
%                       Bibliography                          		   			%
%%%%%%%%%%%%%%%%%%%%%%%%%%%%%%%%%%%%%%%%%%%%%%%%%%%%%%%%%%%%%%%%%%%%%%%%%%%%%%%%%


\end{document}